\documentclass[twocolumn,english,aps,pre,superscriptaddress,showpacs]{revtex4}

\usepackage{amssymb,amsfonts,amsmath}

\usepackage[normalem]{ulem}
\usepackage{pdfpages}
\usepackage{float}

\begin{document}

\title{Evolutionary dynamics of selfish DNA generates pseudo-linguistic features of genomes}

\author{Michael Sheinman$^{1,2}$, Anna Ramisch$^{1}$, Florian Massip$^{1,3}$ and Peter F. Arndt$^{}$}

\address{$^{}$Max Planck Institute for Molecular Genetics, Berlin, Germany\\
$^{2}$Department of Biology, Faculty of Science, Utrecht University, Utrecht, the Netherlands\\
$^{3}$INRA, UR1404 Math\'ematique Informatique Appliqu\'ees du G\'enome \`a l'Environnement-F-78350 Jouy-en Josas, France}



\begin{abstract}
Since the sequencing of large genomes, many statistical features of their sequences have been found. One intriguing feature is that certain subsequences are much more abundant than others. In fact, abundances of subsequences of a given length are distributed with a scale-free power-law tail, resembling properties of human texts, such as the Zipf's law. Despite recent efforts, the understanding of this phenomenon is still lacking.  Here we find that selfish DNA elements, such as those belonging to the Alu family of repeats, dominate the power-law tail. Interestingly, for the Alu elements the power-law exponent increases with the length of the considered subsequences. Motivated by these observations, we develop a model of selfish DNA expansion. The predictions of this model qualitatively and quantitatively agree with the empirical observations. This allows us to estimate parameters for the process of selfish DNA spreading in a genome during its evolution. The obtained results shed light on how evolution of selfish DNA elements shapes non-trivial statistical properties of genomes.
\end{abstract}
\maketitle

\begin{figure}[h]
\centering
\includegraphics[width=0.5\textwidth]{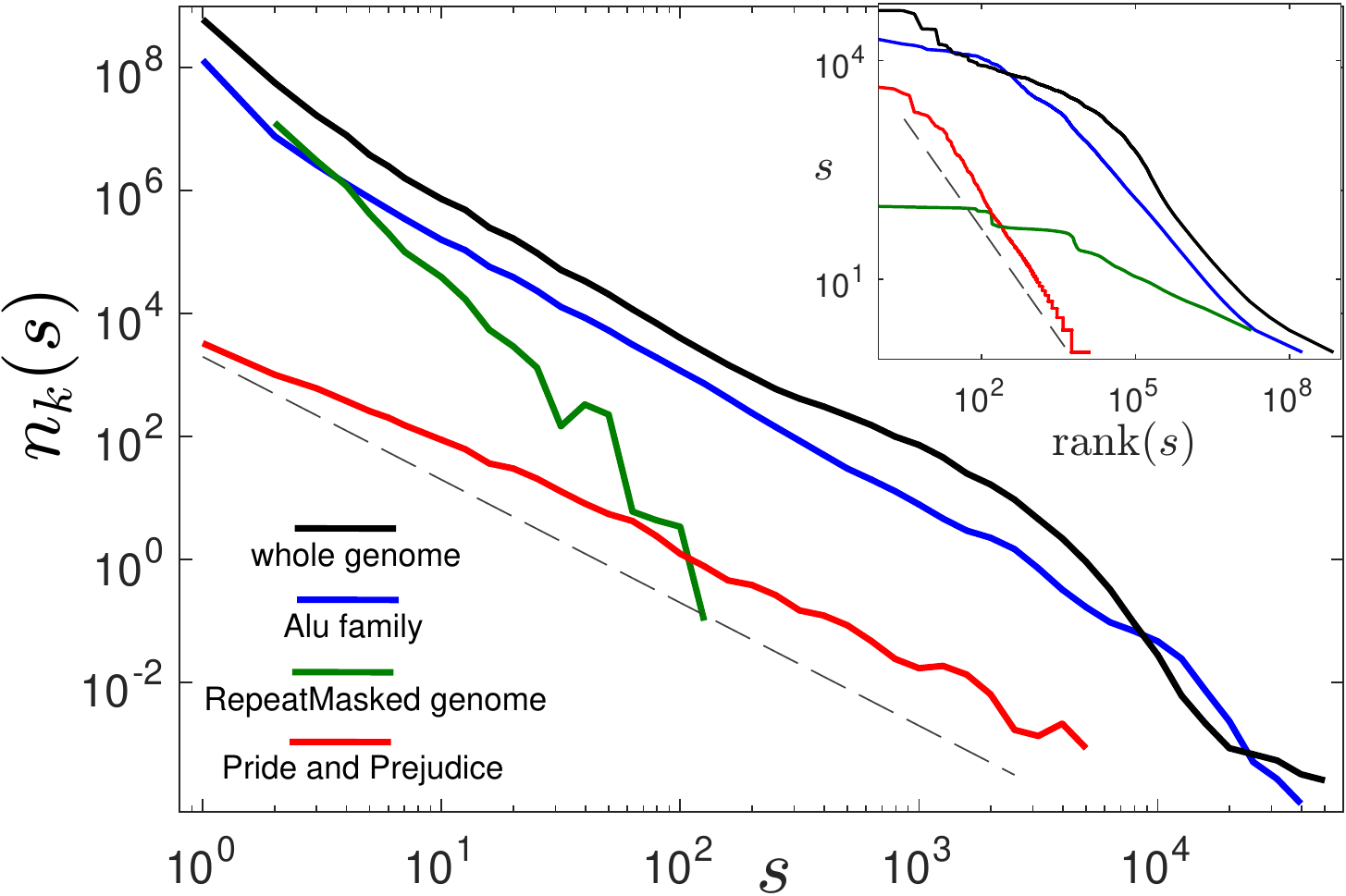}
\caption{Distributions of abundances of $k$-mers for $k=40$. $s$ is the number of copies of a certain $k$-mer and $n_k(s)$ is the number of different $k$-mers with abundance $s$. Distribution for different genomic compartments are presented: the whole genome (solid, black), the whole genome after masking the repeat elements (solid, green) and the Alu family of repeats (solid, blue). See Appendix \ref{Data} for details. For comparison the distribution of word abundances in Pride and Prejudice~\cite{austen1813pride} is also shown (solid, red). The dashed line represents the power-law $n_k(s) \sim s^{-\alpha}$ with $\alpha=2$. For a randomly shuffled human genome or a random sequence of the same length there is not a single $k$-mer with $s>1$. Inset: the corresponding Zipf's plots for the main figure. For each $k$-mer (or a word for Pride and Prejudice) its abundance is plotted vs. the rank of its abundance. The dashed line represents the power-law with an exponent $\alpha=1$.}
\label{Zipf}
\end{figure}

Our genome is a sequence of A, C, G and T nucleotides and can be viewed as a long text of about three billion letters. Only a small part of our genome is functional and under selection~\cite{ohno1972so,ponting2011fraction,kellis2014defining}; the rest (so-called junk DNA) mostly evolves neutrally and, therefore, is naively expected to be a random sequence.
However, the junk DNA contains many homologous sequences, sharing significant similarities to each other. Hence, its statistical properties differ from those of random sequences~\cite{peng1992long,gao2011algebraic,massip2013neutral,massip2015evolution}. One of these properties, which we discuss here, is that for a given length, certain subsequences are much more abundant than others~\cite{mantegna1994linguistic,hsieh2003minimal,sindi2008duplication,chor2009genomic}. Namely, the abundances of $k$-mers---sequences of length $k$---possess a wide, scale-free distribution, as shown in Fig.~\ref{Zipf}. One can see that even for large values of $k$, one finds $k$-mers which appear more than $10^4$ times in the human genome, while in a randomly shuffled genome such $k$-mers would be unique.

This phenomenon resembles statistical properties of human texts, where abundances of words also exhibit a scale-free distribution~\cite{estoup1916gammes}. For human texts such a linguistic feature is often presented as Zipf's~\cite{zipf1949human,newman2005power} or Heaps'~\cite{heaps1978information} law. We exemplify the similarity between the statistics of $k$-mers in the human genome and the statistics of words in human texts in Fig.~\ref{Zipf}. Despite an incomplete analogy, caused by the lack of a natural definition of a word in the genomic context, this intriguing similarity between the genome and human texts has led some researchers to analyze genetic sequences from a linguistic perspective (see, e.g., Refs.~\cite{gimona2006protein,loose2006linguistic}). Other studies called linguistic properties of genomes into question~\cite{israeloff1996can,bonhoeffer1996no,attard1996language,tsonis1997dna,csHuros2007reconsidering}. Here we present an explanation for the observed genomic phenomenon, showing that the pseudo-linguistic features of the $k$-mer abundances statistics in genomes are a consequence of selfish DNA expansion in our genome during its evolution. We develop a model, which accurately reproduces statistical properties of very abundant subsequences in the genome. The model is based solely on selfish spreading of DNA repeats, demonstrating that high abundances of certain $k$-mers do not reflect their functionality for the organism.

Considering different compartments of the genome, one finds that the scale-free distribution of abundances is dominated by subsequences of selfish repetitive DNA elements (see Fig.~\ref{Zipf}). This suggests that the scale-free distribution of abundances is a consequence of the evolutionary dynamics of such elements. Selfish DNA elements (or repeats) are parasitic sequences that duplicate with the help of the cellular machinery of the host organism~\cite{doolittle1980selfish,orgel1980selfish}. Such duplications significantly increase the size of genomes during their evolution and often appear in bursts of activity during a few tens of millions of years~\cite{deininger2002mammalian,batzer2002alu}. After such a burst, the duplication activity stops, but the existing repetitive elements remain in the genome. Some elements acquire a function~\cite{mighell1997alu} or cause a disease~\cite{deininger1999alu}, but most fade away neutrally into the genomic background due to mutations~\cite{brookfield2001selection}. One of the largest and most studied families of repeats in primates is the Alu family, covering $15\%$ of the human genome with more than a million copies~\cite{schmid1996alu,lander2001initial}. In the following, we use the Alu family as a model system to study statistical properties of selfish DNA sequences.

To gain insight into the origin of the observed fat-tailed distributions of $k$-mer abundances in the human genome, we plot in Fig.~\ref{DistNew} the distributions for the Alu family elements for different values of $k$ (see Appendix \ref{Estimating parameters and fitting procedures} for details). One can see that the even the abundances of short $k$-mers are much more dispersed than in a random sequence. For large values of $k>20$ the distributions possess a power-law tail, i.e. $n_k(s) \sim s^{-\alpha}$. Importantly, the exponent of the power-law distribution, $\alpha(k)$, depends on $k$, such that it  increases with $k$, starting from about $2$ for small values of $k$.
This dependence is clearly visible in Fig.~\ref{alphavsk}, where we measure the values of $\alpha(k)$ using the Hill estimator (see Appendix \ref{Estimating parameters and fitting procedures}). A model for the evolutionary dynamics of selfish DNA ought to be able, in particular, to explain these properties of the power-law exponent and, in general, to reproduce the empirical distributions.

In this paper we present a simple model for the evolution of selfish DNA, which accounts qualitatively and quantitatively for the observed distributions of $k$-mer abundances. Using our model, we estimate key parameters of the spreading dynamics of Alu repeat elements and compare them to previous estimates. Our results demonstrate that some non-trivial properties of genomic texts can be understood considering the evolution of selfish DNA, without referring to any linguistic structure of genomes.

\section{The model}
We analyze the following model for the evolution of selfish DNA in a genome. The process starts from the appearance of a single active (i.e. able to duplicate) selfish element of length $L$ at time $t=0$. During the burst of activity, in the time interval $0 \leq t \leq T_1$, all existing \emph{active} elements duplicate in a genome with rate $\gamma$. Each duplication results in a new identical element, which we assume to be \emph{active} with probability $ \delta $ and silent (non-duplicating) with probability $1-\delta$. This results in an exponential growth, such that the average number of elements after the burst of activity ends at time $t=T_1$ is given by 
\begin{equation}
\mathcal{N}=1+\frac{e^{\delta \gamma T_1}-1}{\delta}.
\label{N}
\end{equation} 

\begin{figure}[t]
	\centering
	\includegraphics[width=0.5\textwidth]{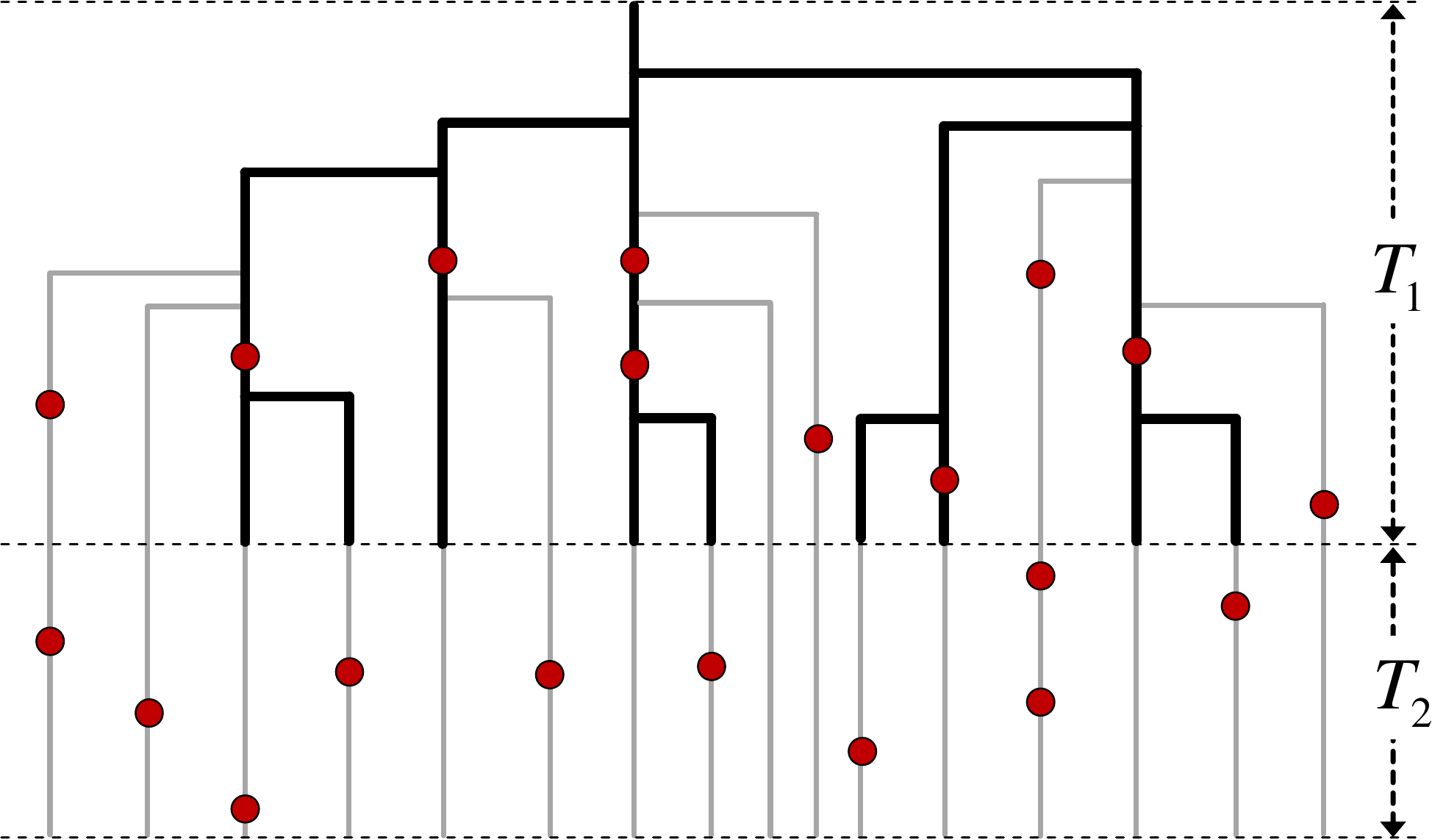}
	\caption{Illustration of the analysed model for the dynamics of repeat elements. Each branch represents a sequence of the repeat. Active elements are depicted in thick, black lines, while silent ones are shown in thin, gray lines. During the activity burst, selfish elements duplicate exponentially with time and accumulate mutations (red marks). After the burst sequences do not duplicate anymore but still mutate.}
	\label{Illustration}
\end{figure}

During such a burst, these duplications shape a branching process, which gives rise to a phylogenetic tree, as illustrated in Fig.~\ref{Illustration}. After the burst, the duplication activity is suppressed and all $\mathcal{N}$ elements are silenced for a time period $T_2$. We observe these elements in the present day genome, at time $t=T_1+T_2$. During and after the burst, all existing elements accumulate mutations with rate $\mu_0$ per bp, except CpG nucleotides, which mutate approximately 6 times faster~\cite{xing2004alu}. We define the effective mutation rate $\mu$ as the weighted average of the two rates. An illustration of the described model for the evolution of selfish DNA elements is presented in Fig.~\ref{Illustration}.

To address the empirically observed scale-free distribution of $k$-mer abundances in genomic data, we consider in particular the statistics of $k$-mers in this model. There are $L-k+1$ $k$-mers in a single element. A duplication event increases the number of all $k$-mers in the duplicating element, while mutations decrease abundances of certain $k$-mers and increase abundances of others. The mutation rate of a $k$-mer is $\mu k$, such that the probability that a $k$-mer does not mutate for a time $T_2$ is given by $e^{-\mu T_2 k}$.

Using two simplifying assumptions, we solve the model analytically (see detailed derivation in the Appendix \ref{Analytic model and its solution}). 
The analytic solution of the model yields that the number of $k$-mers with abundance $s \gg 1$ at present time $t=T_1+T_2$, which we denote by $n_k(s)$, is given by
\begin{equation} 
n_k(s) \simeq (\alpha-2)(L-k+1) \mathcal{N}   \frac{p^{\alpha-1}}{s^{\alpha}}.
\label{dist}
\end{equation}
Here $\mathcal{N}$ is the number of repeat elements at present time, given by Eq.~\eqref{N}. The power-law exponent of the distribution is
\begin{equation}
\alpha(k)=1+\frac{1}{1-\frac{\mu }{\delta\gamma}k}
\label{alpha}
\end{equation}
and $p=e^{-\mu T_2 k}$ is the probability of a $k$-mer to preserve its sequence during the second, silent phase.

Note, that the power-law tail exists only if, on average, a $k$-mer duplicates faster than it mutates, such that  $ \mu k<\delta\gamma$. In the context of this paper this condition is fulfilled. 
The derived dependence for the power-law exponent $\alpha$ in Eq.~\eqref{alpha} accounts for the observations presented above: $\alpha(k)$ is predicted to increase with $k$, starting from $\alpha(k)=2$ for small values of $k$.

To further quantitatively test the presented model, one needs to estimate the parameters $\mathcal{N}$ (number of elements), $\mu$ (effective mutation rate), $T_1$ (time of the first, active phase), $T_2$ (time of the second, silent phase) and $\delta$ (probability of a new element to be active). The duplication rate, $\gamma$ can be then estimated using Eq.~\eqref{N}. We obtain the estimates using the empirical data and the analytic result \eqref{alpha}. As we show below, our model accurately reproduces the empirical data for the Alu family of repeats for the estimated set of parameters.

\begin{figure}[t]
	\centering
	\includegraphics[width=0.5\textwidth]{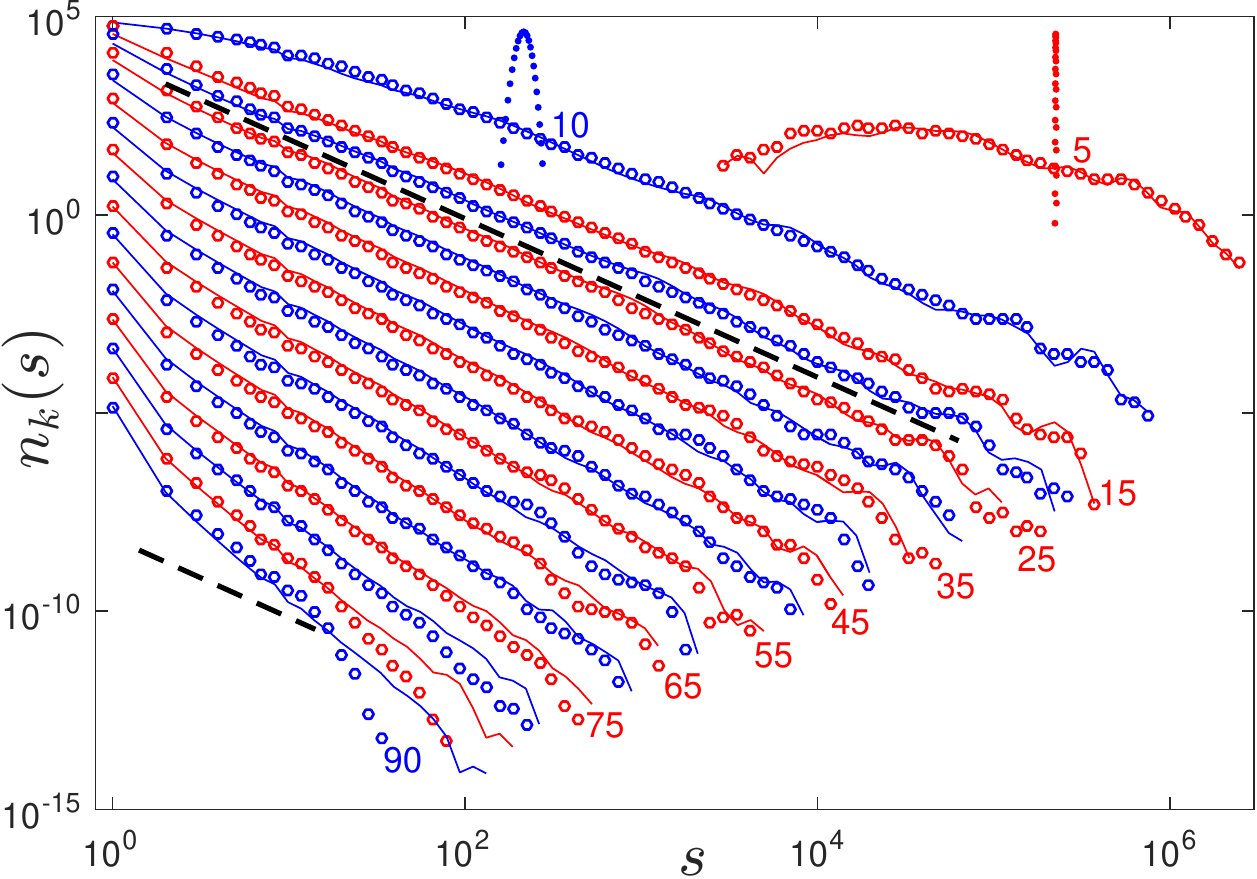}
	\caption{Distributions of abundances of $k$-mers, $n_k(s)$, for different values of $k$, from $5$ to $90$ in steps of $5$, from top to bottom (see numbers in the figure). Circles represent $n_k(s)$ in the empirical data for the Alu family of repeats (see Appendix \ref{Data}). Dots represent $n_k(s)$ in a random sequence, of the same length as the empirical one for $k=5$ (red) and $k=10$ (blue). Lines represent $n_k(s)$ in simulated Alu elements using the set of parameters in Eq.~\eqref{SetParam1} in Appendix \ref{Estimating parameters and fitting procedures}. The results of the simulations do not change drastically, as long as the parameters are within the ranges specified in Eq.~\eqref{SetParam}. The dashed lines represent the power-law decay $n_s \sim s^{-\alpha}$ with $\alpha=2$. For visibility the values of $n_k(s)$ are normalized differently for each value of $k$ (but in the same way for the empirical and the simulated data), so that the units of the vertical axis are arbitrary.}
	\label{DistNew}
\end{figure}
 
\begin{figure}[t]
	\centering
	\includegraphics[width=0.5\textwidth]{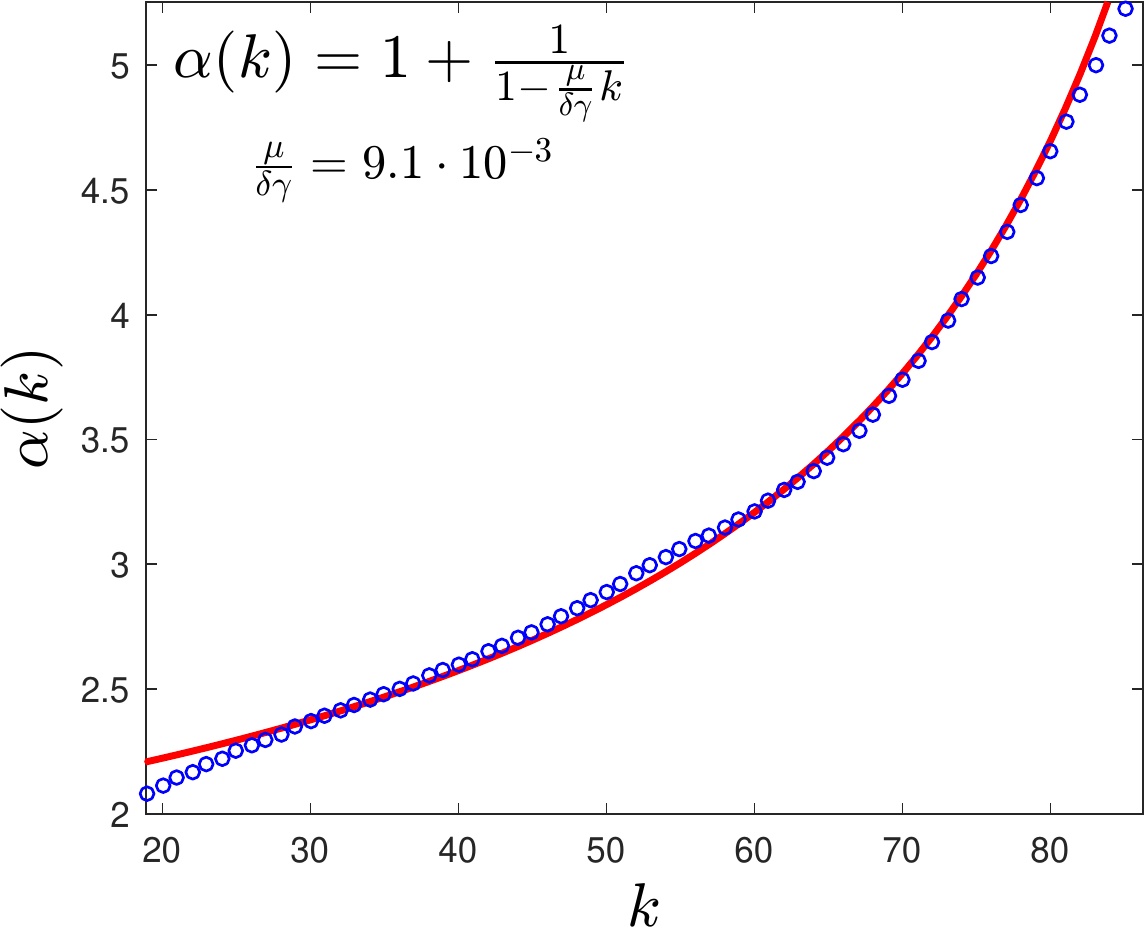}
	\caption{Estimation of parameters of the model using the analytic fit of the empirical data. Circles represent the empirical power-law exponent $\alpha$ as a function of $k$. The line is the numerical fit of the data points using Eq.~\eqref{alpha}, resulting in Eq.~\eqref{est1}. The analytic equation for the fit and the resulting estimator are presented in the upper-left corner. For details of the estimators and the fit see Appendix \ref{Estimating parameters and fitting procedures}.}
	\label{alphavsk}
\end{figure}

\section{Modeling evolution of Alu repeats}
The presented model can be used to study the evolution of large selfish DNA families. We apply it here to the Alu family of repeats, studying distributions of $k$-mer abundances in all identified Alu repeats in the human genome, excluding the still active AluY subfamily~\cite{konkel2015sequence} (see Appendix \ref{Data} for more details). In Fig.~\ref{DistNew} one can see that these distributions qualitatively agree with the predictions  of Eqs.~\eqref{dist} and \eqref{alpha}: the tails of the distributions follow a power-law, the exponents of these power-laws are larger than two and grow with $k$ (see also Fig.~\ref{alphavsk}).
 
We start now with the estimation of the parameters of the model. First, we estimate the ratio $\frac{\mu}{\delta\gamma}$ using the analytic result~\eqref{alpha}.
From the empirical data the value of the power-law exponent $\alpha(k)$ can be estimated for each $k$ using the Hill maximum-likelihood estimator~\cite{arnold1985pareto} (see Appendix \ref{Estimating parameters and fitting procedures}). In Fig.~\ref{alphavsk} one can see the agreement of Eq.~\eqref{alpha} with the empirical power-law exponents, using a fit with a single free parameter, $\frac{\mu}{\delta \gamma}$. The fitting results in 
\begin{equation}
\frac{\mu}{\delta\gamma}= 9.1 \cdot 10^{-3}.
\label{est1}
\end{equation}

To estimate the effective mutation rate, $\mu$, it is important to consider hypermutable CpG di-nucleotides along the Alu elements. In fact, $\mu$ is the average mutation rate of CpG di-nucleotides and other nucleotides.
There are $24$ CpG di-nucleotides in a typical Alu element (e.g. in the consensus sequence of the AluSx subfamily) of length $L \simeq 300$. These di-nucleotides mutate about 6 times faster than other nucleotides on both positions~\cite{xing2004alu}. Thus,
\begin{equation}
\mu \simeq \mu_0\frac{252}{300}+\mu_{\text{CpG}
}\frac{48}{300}=1.8\mu_0,
\label{mu0}
\end{equation} 
where $\mu_0$ is the mutation rate of the non-CpG nucleotides and $\mu_{\text{CpG}}=6\mu_0$ is the mutation rate of the CpG nucleotides. In the following we measure all the rates in units of $\mu_0$, which is of the order of $10^{-9}$yr$^{-1}$, and times in units of $\mu_0^{-1}$. The value of $\mu_0$ is just a global time scale and does not affect the $k$-mer abundances. Nevertheless, in the Discussion section we estimate $\mu_0$, convert all estimated parameters to standard units and compare them with previous estimates in the literature.

The probability of a new element to be active, $\delta$, does not affect the results in the asymptotic limit of large number of Alu elements, $\mathcal{N}$, as long as the effective duplication rate, $\delta \gamma$ is kept constant. However, for a finite value of $\mathcal{N}$ the results change if $\delta$ is too small. In this case estimates of $\alpha(k)$ would be biased to higher values due to highly abundant copies of several active elements, such that Eq.~\eqref{alpha} would not fit well the biased estimates.
The fact that Eq.~\eqref{alpha} does fit well the empirical data indicates that $\delta$ is not very small. Our simulations, with $\mathcal{N}=776710$ and Eq.~\eqref{est1}, indicate that the distribution of abundances does not depend significantly on $\delta$ and Eq.~\eqref{alpha} fits well the data, as long as $\delta$ is above $5\%$. This result supports an earlier study, where $\delta$ is estimated to be $10-20\%$~\cite{cordaux2004retrotransposition}. Using our estimate $\delta=(5-100) \%$ and Eqs.~\eqref{est1},\eqref{mu0} we conclude that the duplication rate is in the range $\gamma=(0.2-4) \cdot 10^{3} \mu_0$. Furthermore, using the above estimates together with Eq.~\eqref{N} we get the estimate $T_1=(5.3-6.9) \cdot 10^{-2}/\mu_0$.

To find the only remaining missing parameter $T_2$, we use the independence of the results on the value of $\delta$ in the relevant regime, setting $\delta=1$ and simulating the model for many different values of $T_2$. The best agreement between the empirical distribution of abundances and the simulated one was obtained for
\begin{equation}
 T_2 = (2.3 \pm 0.1)  \cdot 10^{-2} /\mu_0.
\label{est2}
\end{equation} 
More details about the estimation of the parameters from the empirical data can be found in Appendix \ref{Estimating parameters and fitting procedures}.

In summary, the estimated parameter set for the Alu family evolution model is
\begin{align}
\delta&=(5-100) \% \nonumber\\ 
\gamma&=(0.2-4) \cdot 10^{3} \mu_0\nonumber\\
T_1&=(5.3-6.9) \cdot 10^{-2}/\mu_0\nonumber\\
T_2&=(2.2-2.4) \cdot 10^{-2}/ \mu_0.
\label{SetParam}
\end{align} 
As shown in Fig.~\ref{DistNew}, the model with this set of parameters accurately reproduces the empirical distributions of the $k$-mer abundances for the Alu elements.

\section{Discussion}
There are a few important things to note before we draw conclusions and summarize. First, the presented model is similar in spirit to the one suggested in Ref.~\cite{sindi2008duplication}. However, the basic assumption there was that the evolution of the selfish DNA elements approaches a steady state with a constant genome size, such that any new element replaces an old one, resulting in $\dot{n}_k(s)=0$. As has been shown in Ref.~\cite{sindi2008duplication}, this assumption can only result in an abundance distribution following $n_k(s) \sim s^{-1}$, such that $\alpha=1$ for all values of $k$. In contrast, we assume an exponentially growing steady state of the genome in the burst phase, $\dot{n}_k(s)=\delta \gamma n_k(s)$ (see Eq.~\eqref{SteadyState}). The last assumption makes more biological sense for the expansion of selfish DNA, with a weak or no selection against it. Only in this case, when there is a phase of exponentially expanding repeats, one can get a power-law exponent $\alpha(k)$ which is always larger than $2$ and depends on $k$, as it is observed for the empirical data.

In our model we assumed that CpG di-nucleotides mutate 6 times faster than other nucleotides. This assumption results in an effective mutation rate of Alu elements which is $1.8$ times higher than the mutation rate of non-CpG nucleotides elsewhere in the genome (see Eq.~\eqref{mu0}). A tempting simplification of the model would be to ignore the CpG di-nucleotides, assuming an effective mutation rate for all nucleotides along an Alu element. However, in that case, the distribution obtained for $k$-mer abundances is qualitatively different from the empirical distribution (see Appendix \ref{Importance of CpG di-nucleotides} for more details). Here, we only stress that modeling non-uniform mutation rate with highly mutable CpG di-nucleotides is essential to account for the empirical data.

Evolution of repeat elements in our genome is a complex process, which probably involves selection, population dynamics and other factors~\cite{prak2000mobile,deininger2002mammalian,deininger2003mobile,hedges2004differential,kazazian2004mobile}. Detailed studies of Alu repeats reveal a complex history with many subfamilies appearing at different times~\cite{slagel1987clustering,willard1987existence,jurka1991reconstruction,kapitonov1996age,batzer2002alu,price2004whole,churakov2010novel}. As it often happens in nature, very complex phenomena tend to exhibit random-like statistical features. For instance, complex speciation processes result in a simple Yule statistics of genera sizes~\cite{willis1922some} and simple statistics of pairwise genomic distances~\cite{sheinman2015statistical}; complex biochemical processes result, to some extent, in simple molecular clocks with effective mutation rates of nucleotides and amino acids on the evolutionary timescale~\cite{zuckerkandl1962molecular,kumar2005molecular}, etc. This study suggests another example of this kind: a complex evolution of selfish DNA elements exhibits random-like properties with some effective parameters.

Our estimates of those effective parameters might suffer from various biases. The first stems from the fact that we assumed a constant mutation rate along the human lineage since the origin of the Alu family in the genome, which might have varied, for instance due to different generation times~\cite{li1987evaluation,scally2012revising}. Moreover, Alu elements are enriched (relative to the whole genome) with CpG di-nucleotides which possess an order of magnitude higher mutation rate~\cite{labuda1989sequence,batzer1990structure,xing2004alu}. We assumed that the mutation rate of the CpG di-nucleotides is 6 times higher than that of other nucleotides, but in reality the situation might be more complex~\cite{xing2004alu}. Positive or negative selection can increase or decrease the estimate for the effective mutation rate. The duplication rate can possess more complex temporal structure than the one we assumed in our model and may depend on the sequence of the element~\cite{batzer2002alu}. These and other effects are, probably, the reason for the disagreement between our simulations and the empirical distributions at small abundances and, for very long $k$-mers, at the very end of their tails (see Fig.~\ref{dist}).

Our estimate for the age of the Alu family in units of the neutral mutation rate is $T_1+T_2=(7.7-9.3)\cdot 10^{-2} /\mu_0$. Alternatively, one can estimate the age of the Alu family from the following phylogenetic arguments. The Alu family is primate specific~\cite{liu2009comparative}, so we expect that the age of Alu is about $80 \cdot 10^6$yr~\cite{hedges2015tree}. Therefore, our estimate for the neutral mutation rate turns out to be about $\mu_0=(0.9-1.1) \cdot 10^{-9}$yr$^{-1}$, within the range estimated in the literature~\cite{scally2012revising}.
Furthermore, in this case our estimate of $T_2=(20-24) \cdot 10^6$yr is in a rough agreement with the conclusion of Ref.~\cite{britten1994evidence} that "most human Alu sequences were inserted in a process that ceased about $30$ million years ago". A similar estimate was obtained in Ref.~\cite{shen1991evolution}. 
Therefore, our estimates of the parameters yield reasonable values, suggesting that the above discussed possible biases are not of great importance in the context of this study.

In the literature there are two alternative models for the expansion of Alu elements. The first one is the transposon model, which assumes that every Alu element duplicates with the same rate~\cite{matera1990recently}. This corresponds to $\delta=1$ in our model. The second one, the master gene model, assumes $\delta=0$, implying that there is a single active, duplication potent element which gives rise to all other elements~\cite{shen1991evolution,deininger1992master}. More recent studies suggest that the fraction of active Alu elements, $\delta$, is not 100\% as in the transposon model nor extremely small as in the master gene model~\cite{clough1996computer,johnson2006test,brookfield2006evolution}. 
This fraction for the Alu family was estimated to be $10-20\%$~\cite{cordaux2004retrotransposition}. From our simulations we find that Eq.~\eqref{alpha} is expected to fit the data well as long as $\delta$ is larger than $5\%$. Therefore, the fact that the empirical data is well fitted by Eq.~\eqref{alpha} (see Fig.~\ref{alphavsk}) supports the estimate in Ref.~\cite{cordaux2004retrotransposition}.

Since the assumptions of our model are quite general, it can capture
the dynamics of evolution of other selfish elements. For instance,
qualitatively similar results can be also observed for the L1
family of repeats. Selfish elements cover a significant fraction
of our genome, resulting in a genome-wide power-law distribution of
$k$-mer abundances. Since different selfish DNA families evolve with different effective parameters, their mixture---the genome-wide power-law --- is not expected to be clean. However, the main
predicted feature of our model is that the power-law exponent has to be larger than and
close to 2, as it is observed. Thus, our results explain the power-law in
the Zipf plot of $k$-mers in genomes, without referring to any
linguistic or functional features. In fact, we demonstrate that a high abundance of a certain subsequence is not necessarily due to its functionality for an organism, but may rather reflect its ability to parasitize and selfishly spread in the host genome. The presented simple model of selfish DNA evolution surprisingly accurately accounts for statistical properties of these highly abundant subsequences in our genome.

\begin{acknowledgments}
We thank M.A. Batzer, A. Fratzl, A. Mammana and R. Hermsen for fruitful discussions.
\end{acknowledgments}

\appendix

\section{Data} 
\label{Data}
The sequences of all identified Alu repeat elements in the human genome were downloaded from the Ensembl database using the Perl API~\cite{cunningham2015ensembl}. We filter out the X and Y chromosomes and the sequences belonging to the still active AluY subfamily~\cite{konkel2015sequence}. In Fig.~\ref{Zipf} the AluY subfamily is not filtered out. The $k$-mer abundances were counted using the Jellyfish program~\cite{marccais2011fast}. To smooth the resulting distribution of abundances in Figs.~\ref{dist} and \ref{DistCpG}, we used logarithmic binning for $s>7$, such that the ratio between two neighboring values of $s$ is $1.1885$.
 
\section{Simulations} \label{Simulations} 
First we computed a branching pattern (or phylogenetic tree) of selfish elements as shown schematically in Fig.~\ref{Illustration}. This branching process was simulated in continuous time using a Kinetic Monte Carlo scheme~\cite{bortz1975new}. We start with one active element at $t=0$. At any given point in time all possible future branchings of active elements are listed; each of them occur with rate $\gamma$. Drawing a random number one of them is selected and executed; the time then advanced appropriately.
The new element is active, i.e. able to duplicate again, with probability $\delta$. Drawing another random number we determine whether the new edge is active or not. When the total number of elements approaches the empirical one $\mathcal{N} = 776710$ we rescale the length of all edges, such that the height of the tree is $T_1$. After this the terminal edges of the tree are extended by the time $T_2$, such that the height of the tree is $T_1+T_2$. 

Once the phylogenetic tree is computed, we simulate the evolution of nucleotide sequences along its edges. At the root we start with the ancestral AluSx master sequence, which is mutated along the edges and duplicated at the nodes of the phylogenetic tree. Mutations are again modeled by Kinetic Monte Carlo. Nucleotides change to one of the other 3 nucleotides with rate $\mu_0$. To model the hyper-mutability of CpGs, we allow Cs and Gs in CpGs to change to T or A, respectively, with an increased rate $\mu_{\rm CpG}=6\mu_0$.

\section{Estimating parameters and fitting procedures}\label{Estimating parameters and fitting procedures}
 The number of Alu elements in the empirical data we estimate as the average of $\sum_{s=1}^\infty s n_k/(L-k+1)$ over all values of $k$ in the range $5 \leq k\leq90$, with $L=300$. This results in $\mathcal{N} = 776710$.

Using the empirical data we estimate the value of the power-law exponent of the $k$-mer abundances distributions tail, $\alpha(k)$, using the Hill maximum-likelihood discrete estimator~\cite{arnold1985pareto,clauset2009power} for $ s \geq 3$. Namely, for each $k$,
\begin{equation}
	\alpha(k)=1+\frac{\sum_{s=3}^\infty n_k(s)}{\sum_{s=3}^\infty n_k(s) \ln \frac{s}{3-\frac{1}{2}}}.
\end{equation}

The obtained values of $\alpha(k)$ are fitted in the range $35\leq k \leq 75$ with Eq.~\eqref{alpha} with $\frac{\mu}{\delta\gamma}$ as a single free parameter [see Fig.~\ref{alphavsk}] using the Levenberg-Marquardt nonlinear least squares algorithm~\cite{george2003nonlinear} in Matlab. This way we get the estimate~\eqref{est1}.

The value of $T_2$ we fit manually by simulating the model with many different values of $T_2$ with $\delta=1$. This results in estimate~\eqref{est2}. With estimates \eqref{est1} and \eqref{est2} we simulate the model changing the value of $\delta$. The results do not change significantly from the $\delta=1$ case, as long as $\delta$ is not below $5\%$. Furthermore, below this threshold one cannot fit the empirical estimates of $\alpha(k)$ with Eq.~\eqref{alpha}. Since the empirical results are well fitted with Eq.~\eqref{alpha}, we conclude that $\delta$ is the range $0.05 \leq \delta \leq 1$. 
The resulting set of estimated parameters used for simulations is
\begin{align}
\delta&=1 \nonumber\\ 
\mathcal{N}&=776710 \nonumber\\
T_1&=6.9 \cdot 10^{-2}/\mu_0\nonumber\\
T_2&=2.3 \cdot 10^{-2}/ \mu_0.
\label{SetParam1}
\end{align} 
The results of the simulations of the model with these parameters can be seen in Fig.~\ref{DistNew}.
For $\delta$ in the estimated range $0.05 \leq \delta \leq 1$ the estimates for the parameters of the model are given in Eq.~\eqref{SetParam}.

\section{Analytic model and its solution}\label{Analytic model and its solution}
To solve the model of selfish DNA evolution analytically we consider two simplifying assumptions:

1. If a mutation happens in a $k$-mer this $k$-mer becomes a new, unique sequence in the genome. While this assumption is valid for large values of $k$, the mutated $k$-mer has a significant chance not to be unique, i.e., to be present elsewhere in the genome, for small values of $k$. That is why the abundance distribution of short $k$-mers is not well described by the analytic model, but agrees with the results of the simulations of the full model without the simplifying assumptions. And this is the reason why we fit the parameters using the analytic solution only for large values of $k \geq 30$.

2. The second assumption is that all elements are active but with the reduced effective duplication rate $\delta \gamma$.

Within this framework, let us consider now $n_k(s,t)$---the average number of different $k$-mers, which appear $s$ times in all copies of the repeat during the burst, at time $0 \leq t \leq T_1$. Starting from a single element the initial condition is given by $n_k(s,t=0)=(L-k+1)\delta_{s,1}$. The dynamic equation for $n_k(s,t)$ for $s>1$ is given by
\begin{align}
\dot n_k(s,t)=&\delta\gamma(s-1)n_k(s-1,t)+ \nonumber \\
&+\mu k  (s+1) n_k(s+1,t)-s(\delta\gamma+\mu k) n_k(s,t).
\label{nk}
\end{align}
The first term is the gain of $n_k(s,t)$ from duplications of $k$-mers which appear $s-1$ times. The second is the gain term from mutations (the effective mutation rate for a $k$-mer is $\mu k$, assuming independently mutating base-pairs) of $k$-mers which appear $s+1$ times. The third term is the loss of $n_k(s,t)$ from duplications or mutations of $k$-mers which appear $s$ times. The dot denotes the time derivative.

Every mutation of a $k$-mer is assumed to generate a unique $k$-mer, with abundance $s=1$. This is reflected in the equation for $n_k(s=1,t)$, which takes the form
\begin{align}
\dot n_k(1,t)=&  \mu k\left[ \mathcal{N}(t)(L-k+1)- n_k(1,t)+2 n_k(2,t)\right]- \nonumber \\
 &-\delta\gamma n_k(1,t),
\label{nk1}
\end{align}
where 
\begin{equation}
\mathcal{N}(t)=1+\frac{e^{\delta \gamma t}-1}{\delta}
\label{NA}
\end{equation} 
is the total number of repeat elements at time $t$. The gain term in Eq.~\eqref{nk1} is due to mutations of all repeat elements (excluding those with abundance $s=1$). Note that a mutation of a $k$-mer with $s=2$ generates two $k$-mers with $s=1$. The loss term is due to duplications of $k$-mers with $s=1$ copies.

The equations  for the dynamics of the abundances distribution during the burst phase, \eqref{nk} and \eqref{nk1}, can be solved analytically to any required precision in the steady state limit, which in this burst phase exhibits an exponential growth of $n_k(s,t)$ with rate $\delta \gamma$, for all finite $s$ values, such that 
\begin{equation}
\dot n_k(s,t)=\delta\gamma n_k(s,t). 
\label{SteadyState}
\end{equation}
In this limit the solution of Eq.~\eqref{nk} for large values of $s$ is given by 
\begin{equation}
n_k(s,t) \simeq  (\alpha-2) (L-k+1)\mathcal{N}(t) \frac{1}{s^\alpha},
\label{asymptot}
\end{equation}
where the power-law exponent $\alpha$ is given by Eq. \eqref{alpha}.
The prefactor is obtained using the normalization condition,
\begin{equation}
	\sum_{s=1}^\infty s n_k(s,t)=(L-k+1)\mathcal{N}(t).
\end{equation}

After the burst ends at time $t=T_1$, abundances of non-unique $k$-mers decrease on average due to mutations. The probability of a $k$-mer to preserve its sequence for time $T_2$ without mutations is $p=e^{-\mu T_2 k}$. Thus, the distribution of abundances for $s>1$ at present time $t=T_1+T_2$ is given by
\begin{equation}
n_k(s,t=T_1+T_2)=\sum_{j=s}^\infty n_k(j,T_1) \binom {j} {s} p^s (1-p)^{j-s}.
\label{binom}
\end{equation}
For $s=1$ the number of $k$-mers is further increased after the burst due to mutations of non-unique $k$-mers and is given by
\begin{align}
&n_k(1,t=T_1+T_2)= \nonumber\\
&\sum_{j=1}^\infty n_k(j,T_1) \binom {j} {1} p^1 (1-p)^{j-1}+ (L-k+1)\mathcal{N} (1-p) .
\label{binom1}
\end{align}
Saddle point approximation of Eq~\eqref{binom} (the saddle point is at $j=s/p \gg1$) results in Eq.~\eqref{dist}.


\section{Importance of CpG di-nucleotides}\label{Importance of CpG di-nucleotides}
\begin{figure*}[h!]
	\centering
	\includegraphics[width=\textwidth]{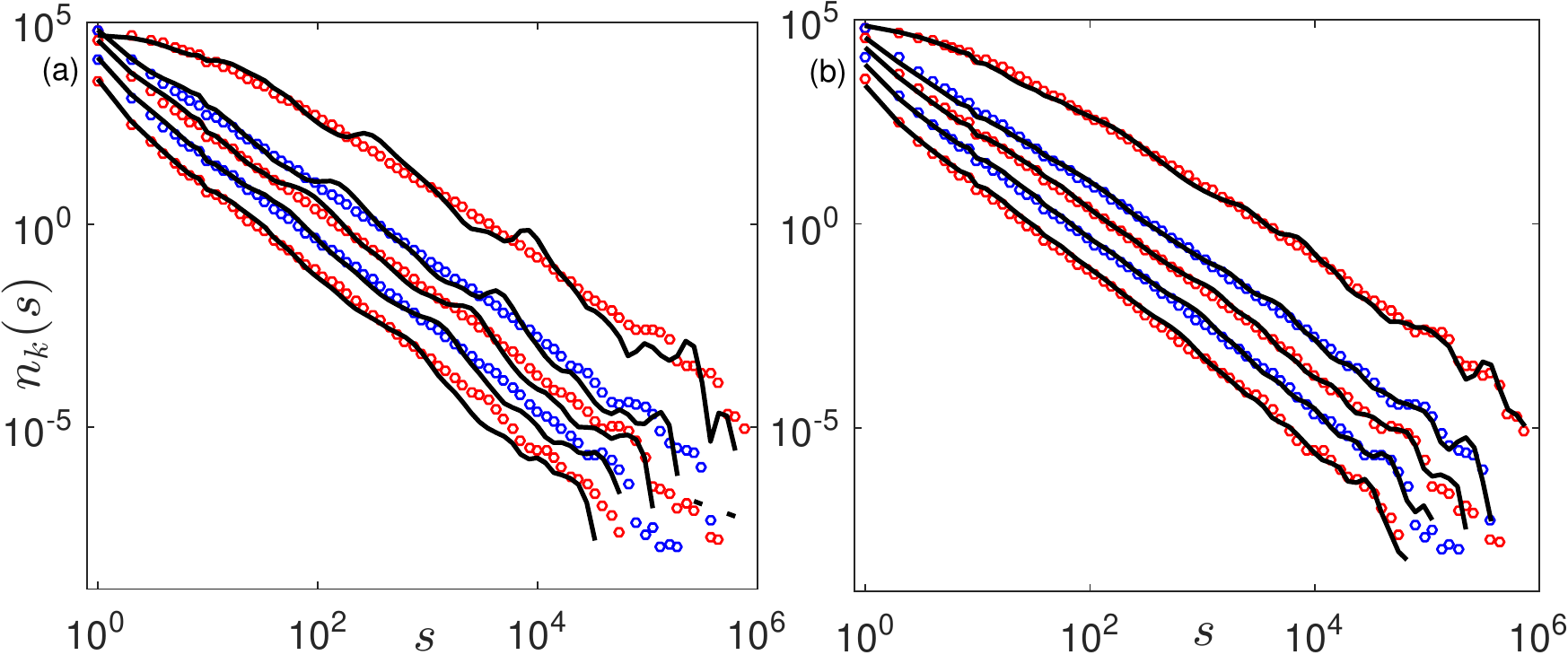}
	\caption{Distributions of abundances of $k$-mers for different values of $k$, from $10$ to $30$ in steps of $5$, from top to bottom. Circles represent abundances of k-mers in the empirical data of Alu elements (see Appendix \ref{Data} for details). Lines represent abundances of k-mers in simulated Alu elements using the set of parameters in Eq.~\eqref{SetParam1} with (a) uniform mutation rate $\mu=1.8 \mu_0$ and with (b) non-uniform mutation rate $\mu_0$ for non-CpG nucleotides and $6\mu_0$ for CpG di-nucleotides. In fact panel (b) is merely a zoom in on Fig.~\ref{DistNew}. For visibility the values of $n_k(s)$ are normalized differently for each value of $k$ (but in the same way for empirical and simulated data), so that the units of the vertical axis are arbitrary.}
	\label{DistCpG}
\end{figure*}

To assess the importance of the non-uniform mutation rate with 6 times more mutable CpG di-nucleotides we performed simulations with a uniform mutation rate equal to the effective one, $1.8  \mu_0$ (see Eq.~\eqref{mu0}).
As shown in Fig.~\ref{DistCpG}, the results of the simulations significantly deviate from simulations with more mutable CpG di-nucleotides and from the empirical results. 
In Fig.~\ref{DistCpG}(a) one can see that although the overall structure of the distribution with uniform mutation rate is similar to the empiric ones, the power-law is not as clean and possesses a rather "bumpy" shape. This is also in contrast to the result of the analytic model, which predicts a clean power-law behaviour in the asymptotic regime $s \gg 1$. Therefore, the reason for the bumps is that, in contrast to the assumption of the analytic model, not every mutation of a $k$-mer leads to a new, unique $k$-mer, but can also generate a $k$-mer which already exists in the genome. If one takes the non-uniform mutation rate into account in the simulations, the "bumps" almost disappear and the resulting distribution is much more similar to the empirical one and the one predicted by the analytic model [see Fig.~\ref{DistCpG}(b) and Fig.~\ref{DistNew}]. 

So, why do CpG di-nucleotides make the power-law cleaner? The reason is as follows. Due to their high mutation rate the divergence of most Alu sequences at the CpG positions is about $36\%$, which is much higher than at the non-CpG positions with about $7\%$~\cite{xing2004alu}. This makes these CpG positions similar to bps with random nucleotides. In fact, if one substitutes bps at CpG positions by random bps the resulting distribution of abundances does not change significantly, being still similar to the empirical and analytic distributions with a clean power-law. Now the question reduces to: why do a few random bps along the selfish elements smooth out the otherwise "bumpy" distribution of abundances? To understand this, let us start with a set of $k$-mers simulated with a uniform mutation rate resulting in the "bumpy" $k$-mer abundances distribution $n_k(s)$, as shown in Fig.~\ref{DistCpG}(a). Consider, for simplicity, adding one random bp (say, with $1/2$ probability of C and T) to each $k$-mer. Then the distribution of abundances of the resulting $k+1$-mers is given by
\begin{equation}
	\bar{n}_{k+1}(s)=2\sum_{q=s}^\infty  \binom{q}{s}\left(\frac{1}{2}\right)^qn_k(q).
	\label{bumps}
\end{equation}
This weighted summation over $n_k(s)$ smooths the distribution, such that the resulting distribution $\bar{n}_{k+1}(s)$ has the same asymptotics, but lacks large "bumps". These qualitative results also hold if there are 4 possible nucleotides with arbitrary fractions, as long as none of the fractions is close to one. If the fraction of one nucleotide is close to one then there is no smoothing due to the summation in Eq.~\eqref{bumps}, but merely an increase of the fraction of unique $k$-mers, like in summations \eqref{binom} and \eqref{binom1} with a small value of $p$. To summarize this part we conclude, that the presence of highly mutable CpG di-nucleotides in Alu elements smooths the abundances distribution, bringing it closer to the predictions of the analytic model and the empirical results. This is why we modeled the evolution of Alu elements taking into account the presence of CpG di-nucleotides resulting in a non-uniform mutation rate along the Alu elements.

\bibliographystyle{apsrev}
\bibliography{citations}
\end{document}